\documentclass[prc,reprint,showpacs,preprintnumbers,amsmath,amssymb,superscriptaddress,floatfix]{revtex4-1}
\usepackage{graphicx}
\usepackage{dcolumn}
\usepackage{bm}
\usepackage{floatrow}
\usepackage[caption=false]{subfig}
\usepackage{booktabs}
\usepackage{multirow}
\usepackage{siunitx}
\usepackage{subfig}
\usepackage{longtable}
\usepackage{footnote}
\usepackage{afterpage}
\usepackage{tablefootnote}
\begin{document}
\preprint{}

\title{Cross-shell excited configurations in the structure of $^{34}$Si}

\author{R.S.~Lubna}
\affiliation{TRIUMF, 4004 Wesbrook Mall, Vancouver, BC, V6T 2A3, Canada}
\affiliation{Facility for Rare Isotope Beams, Michigan State University, East Lansing, Michigan 48824, USA}
\author{A.B.~Garnsworthy}
\affiliation{TRIUMF, 4004 Wesbrook Mall, Vancouver, BC, V6T 2A3, Canada}
\author{Vandana Tripathi}
\affiliation{Department of Physics, Florida State University, Tallahassee, Florida 32306, USA}
\author{G.C. ~Ball}
\affiliation{TRIUMF, 4004 Wesbrook Mall, Vancouver, BC, V6T 2A3, Canada}
\author{C.R.~Natzke}
\affiliation{TRIUMF, 4004 Wesbrook Mall, Vancouver, BC, V6T 2A3, Canada}
\affiliation{Department of Physics, Colorado School of Mines, Golden, CO 80401, USA}
\author{M.~Rocchini}
\altaffiliation{Present address:INFN Sezione di Firenze, IT-50019 Firenze, Italy}
\affiliation{Department of Physics, University of Guelph, Guelph, Ontario N1G 2W1, Canada}
\author{C.~Andreoiu}
\affiliation{Department of Chemistry, Simon Fraser University, Burnaby, British Columbia V5A 1S6, Canada}
\author{S.S.~Bhattacharjee}
\affiliation{TRIUMF, 4004 Wesbrook Mall, Vancouver, BC, V6T 2A3, Canada}
\author{I.~Dillmann}
\affiliation{TRIUMF, 4004 Wesbrook Mall, Vancouver, BC, V6T 2A3, Canada}
\affiliation{Department of Physics and Astronomy, University of Victoria, Victoria, British Columbia V8P 5C2, Canada}
\author{F.H.~Garcia}
\altaffiliation{Present address: Nuclear Science Division, Lawrence Berkeley National Laboratory, Berkeley, California 94720, USA}
\affiliation{Department of Chemistry, Simon Fraser University, Burnaby, British Columbia V5A 1S6, Canada}
\author{S. A.~Gillespie}
\altaffiliation{Present Address: Facility for Rare Isotope Beams, Michigan State University, East Lansing, Michigan 48824, USA}
\affiliation{TRIUMF, 4004 Wesbrook Mall, Vancouver, BC, V6T 2A3, Canada}
\author{G.~Hackman}
\affiliation{TRIUMF, 4004 Wesbrook Mall, Vancouver, BC, V6T 2A3, Canada}
\author{C.J.~Griffin}
\affiliation{TRIUMF, 4004 Wesbrook Mall, Vancouver, BC, V6T 2A3, Canada}
\author{G.~Leckenby}
\affiliation{TRIUMF, 4004 Wesbrook Mall, Vancouver, BC, V6T 2A3, Canada}
\affiliation{Department of Physics and Astronomy, University of British Columbia, Vancouver, BC V6T 1Z4, Canada}
\author{T.~Miyagi}
\affiliation{TRIUMF, 4004 Wesbrook Mall, Vancouver, BC, V6T 2A3, Canada}
\affiliation{Technische Universit\"at Darmstadt, Department of Physics, 64289 Darmstadt, Germany}
\affiliation{ExtreMe Matter Institute EMMI, GSI Helmholtzzentrum f\"ur Schwerionenforschung GmbH, 64291 Darmstadt, Germany}
\affiliation{Max-Planck-Institut f\"ur Kernphysik, Saupfercheckweg 1, 69117 Heidelberg, Germany}
\author{B.~Olaizola}
\altaffiliation{Present address: Instituto de Estructura de la Materia, CSIC, 28006 Madrid, Spain}
\affiliation{TRIUMF, 4004 Wesbrook Mall, Vancouver, BC, V6T 2A3, Canada}
\author{C.~Porzio}
\altaffiliation{Present address: Nuclear Science Division, Lawrence Berkeley National Laboratory, Berkeley, California 94720, USA}
\affiliation{TRIUMF, 4004 Wesbrook Mall, Vancouver, BC, V6T 2A3, Canada}
\affiliation{L'Istituto Nazionale di Fisica Nucleare (INFN) Sezione of Milano and Dipartimento di Fisica, Universit\`{a} di Milano, I-20133 Milano, Italy}
\author{M.M.~Rajabali}
\affiliation{Department of Physics, Tennessee Technological University, Cookeville, Tennessee, 38505, USA}
\author{Y.~Saito}
\affiliation{TRIUMF, 4004 Wesbrook Mall, Vancouver, BC, V6T 2A3, Canada}   
\affiliation{Department of Physics and Astronomy, University of British Columbia, Vancouver, BC V6T 1Z4, Canada}
\author{P.~Spagnoletti}
\affiliation{Department of Chemistry, Simon Fraser University, Burnaby, British Columbia V5A 1S6, Canada}
\author{S.L. Tabor}
\affiliation{Department of Physics, Florida State University, Tallahassee, Florida 32306, USA}
\author{R.~Umashankar}
\affiliation{TRIUMF, 4004 Wesbrook Mall, Vancouver, BC, V6T 2A3, Canada}
\affiliation{Department of Physics and Astronomy, University of British Columbia, Vancouver, BC V6T 1Z4, Canada}
\author{V.~Vedia}
\affiliation{TRIUMF, 4004 Wesbrook Mall, Vancouver, BC, V6T 2A3, Canada}
\author{A.~Volya}
\affiliation{Department of Physics, Florida State University, Tallahassee, Florida 32306, USA}
\author{J.~Williams}
\affiliation{TRIUMF, 4004 Wesbrook Mall, Vancouver, BC, V6T 2A3, Canada}
\author{D.~Yates}
\affiliation{TRIUMF, 4004 Wesbrook Mall, Vancouver, BC, V6T 2A3, Canada}
\affiliation{Department of Physics and Astronomy, University of British Columbia, Vancouver, BC V6T 1Z4, Canada}

\date{\today}

\begin{abstract}
The cross-shell excited states of $^{34}$Si have been investigated via $\beta$-decays of the $4^-$ ground state and the $1^+$ isomeric state of $^{34}$Al. Since the valence protons and valence neutrons occupy different major shells in the ground state as well as the intruder $1^+$ isomeric state of $^{34}$Al, intruder levels of $^{34}$Si are populated via allowed $\beta$ decays. Spin assignments to such intruder levels of $^{34}$Si were established through $\gamma$-$\gamma$ angular correlation analysis for the negative parity states with dominant configurations $(\nu d_{3/2})^{-1} \otimes (\nu f_{7/2})^{1}$ as well as the positive parity states with dominant configurations  $(\nu sd)^{-2} \otimes (\nu f_{7/2}p_{3/2})^2$. The configurations of such intruder states play crucial roles in our understanding of the $N=20$ shell gap evolution. A configuration interaction model derived from the FSU Hamiltonian was utilized in order to interpret the intruder states in $^{34}$Si. Shell model interaction derived from a more fundamental theory with the Valence Space In Medium Similarity Renormalization Group (VS-IMSRG) method was also employed to interpret the structure of $^{34}$Si. 
\end{abstract}

\maketitle


\section{Introduction}
The nuclei with large proton-to-neutron ratio, centered at $Z \approx 11$ and $N \approx 20$ have provided intriguing evidence of the fragility of the long believed permanent magic number, $N=20$ \cite{thibault, warburton}. The bizarre phenomena of the nuclei with $10 \leq Z \leq 12$ and $N \approx 20$, exhibiting the ground states dominated by intruder configurations, have categorized them as members of the Island of Inversion (IoI) in the nuclear landscape. The observation of the exotic phenomena in the nuclear structures has triggered numerous experimental and theoretical efforts to scrutinize the persistence of the canonical shell gaps. A number of experiments have been performed around $N = 20$ along with the development of the sophisticated theoretical models in order to understand the structural evolution associated with the increasing ratio of $N/Z$ that leads from stable nuclei to the $N=20$ IoI region \cite{thibault, MOTOBAYASHI, huber, detraz, GUILLEMAUD, YANAGISAWA, Tripathi}. 

Although the $0^+$ ground state of $^{34}$Si has been suggested to be dominated by the normal configuration, and, hence $^{34}$Si is not explicitly a member of the $N \approx 20$ IoI, it is of significant importance for our understanding on the evolution of the $sd$-$fp$ shell gap. $^{34}$Si is the transitional nucleus along the even-mass $N=20$ isotonic chain with $^{30}$Ne, $^{32}$Mg being the members of the IoI and $^{36}$S, $^{38}$Ar having the normal ground-state configurations along the same chain. The low-lying positive-parity excitations in $^{34}$Si exhibit both normal and intruder configurations. In a study conducted by Rotaru $et\, al.$ \cite{rotaru}, the first excited $0^+$ state at 2719 keV in $^{34}$Si has been identified via the $\beta$ decay of the $1^+$ isomeric state of $^{34}$Al, which was interpreted as having a dominant two particle two hole (2p2h) configuration.  

While the focus has mainly been on the positive parity states in the previous studies, the negative parity intruder states are also quite informative in understanding the structural evolution. In the simplest shell model picture, the low-lying negative-parity states in $^{34}$Si can be constructed by the $(\nu d_{3/2})^{-1} \otimes (\nu f_{7/2})^{1}$ and $(\nu d_{3/2})^{-1} \otimes (\nu p_{3/2})^{1}$ couplings giving rise to the states with spins $2^-$ - $5^-$ and $0^-$ - $3^- $ respectively. These states will mainly be generated by the promotion of one neutron from the $sd$ to the $fp$ shell. The promotion of three particles across the shell gap is expected to cost more energy. Among these negative parity states, $4^-$ and $5^-$ are uniquely constructed from the dominant $(\nu d_{3/2})^{-1} \otimes (\nu f_{7/2})^{1}$ configuration and, hence, sensitive to the $N = 20$ shell gap. States of spin $3^-$ will arise from both couplings and will provide information on the $N=20$ and $28$ shell gaps. Although, the negative parity states were populated in previous experiments \cite{lica34Si, han34Si, nummela34Si}, their spins were all tentatively assigned except for the lowest-lying $3^-$ state. 

In the present work, both the negative and positive parity states of $^{34}$Si have been populated via the $\beta$-decay of $^{34}$Al. The parent nucleus $^{34}$Al has a $4^-$ ground state and a isomeric state of $1^+$ with half-lives 53.73(13) ms and 22.1(2) ms respectively as reported in Ref. \cite{lica34Si}. In the current study, the $\beta$-decay of $^{34}$Al has been utilized as an effective tool to strongly populate specific levels of $^{34}$Si that has allowed an analysis targeting firm spin assignments.

The configuration interaction shell-model method FSU \cite{lubna1, lubna2} and an ab-initio approach with the Valence Space In Medium Similarity Renormalizarion Group (VS-IMSRG) \cite{vsimsrg} method have been utilized to discuss the cross-shell excited configurations of $^{34}$Si.


\section{Experimental Details and Procedures}

The experiment was carried out at the Isotope Separator and Accelerator (ISAC) facility at TRIUMF which utilizes the isotope separation on-line (ISOL) technique to produce rare-isotope beams (RIB). $^{34}$Al ions were produced in the reactions of a 480\,MeV proton beam of 14\,$\mu$A intensity, delivered by the TRIUMF cyclotron impinging on a UCx target with a thickness of 9.89 $g/cm^2$. The TRIUMF Resonant Ionization Laser Ion Source (TRILIS) \cite{trilis} was used in order to selectively ionize $^{34}$Al atoms. The ions were then accelerated to 40\,keV, separated by the ISAC mass separator and delivered to the $\beta$-decay station at the Gamma-Ray Infrastructure For Fundamental Investigations of Nuclei (GRIFFIN) \cite{griffin}. The $^{34}$Al ions were implanted into a mylar tape at the center of the GRIFFIN array at a beam intensity which varied between 160 and 300\,pps. The tape was moved continuously at a speed of 1 $cm/s$ in order to reduce the long-lived descendants from the $^{34}$Al decay chain. The speed was chosen such that five half-lives of the $4^-$ ground state would elapse before the implanted nuclei moved more than 3 mm from the center of the GRIFFIN array to maximize the $^{34}$Si statistics and to reduce the background from other decaying daughters along with minimizing the distortion of the $\gamma$-$\gamma$ angular correlations.  

The de-exciting $\gamma$-rays following $\beta$-decay were detected by the GRIFFIN array, which consisted of 16 Compton-suppressed High-Purity Germanium (HPGe) clover detectors arranged in a rhombicubocatahedral geometry and positioned at 145\,mm from the implantation point. $\beta$ particles were detected by two detection systems; the Scintillating Electron Positron Tagging ARray (SCEPTAR) \cite{griffin} consisting of 10 plastic scintillator paddles arranged in two pentagonal concentric rings which was placed in the upstream side of the decay chamber and a 1\,mm thin, fast plastic scintillator called the Zero Degree Scintillator (ZDS) \cite{griffin} which was located in the downstream side of the beam implantation point. Eight $\text {LaBr}_3\text{(Ce)}$ detectors along with the bismuth germanate suppressors were used in the present experimental setup, however, data from this detection system were not considered in the current analysis. A 20 mm delrin absorber surrounded the implantation chamber which reduced the background from the bremsstrahlung process and suppressed very high energy $\beta$ particles at the same time. A custom built GRIFFIN digital data acquisition (DAQ) system \cite{daq} was used to record energy and timing signals from all detector types in a triggerless mode.

The HPGe events were sorted in the single crystal mode into both $\gamma$ singles and $\gamma-\gamma$ coincidence matrices with and without the $\gamma$ transitions in coincidence with the $\beta$ particles detected at the SCEPTAR and the ZDS. A quadratic energy calibration of the HPGe detectors was performed with the standard sources $^{152}$Eu and $^{56}$Co. A $^{16}$O peak at 6129 keV energy, most likely produced by the inelastic neutron scattering of fast neutrons from the ISAC target, was observed as a background peak and was also used in the energy calibration in order to achieve a better calibration at higher energies. The standard sources $^{133}$Ba, $^{152}$Eu and $^{56}$Co together with a GEANT4 simulation have been utilized for the efficiency calibration of the HPGe detectors. All the observed $\gamma$-ray transitions were corrected for the summing effect, both summing-in and summing-out, following the procedure and the conventions described in Ref. \cite{griffin}.

The resulting data were employed in extending the $^{34}$Si level scheme by using the $\gamma$-$\gamma$ coincidence method. Along with the good statistics, the GRIFFIN array offers a wide angular coverage to perform $\gamma$-$\gamma$ angular correlations in order to determine the multipolarity of the de-exciting $\gamma$-ray transitions, and hence, the spin assignment to the corresponding levels. The angular correlations performed in this work have followed the procedure discussed in Ref. \cite{angcor}. There is a total of 51 individual angular bins for the clover detector crystals associated with the GRIFFIN array.  These 51 bins were grouped into 7 in order to enhance the statistics at each angle. Method 2 described in the aforementioned reference was implemented where the experimental angular correlation data were fitted to the linear combination of different components of Legendre polynomials obtained with the GEANT4 simulations in order to properly account for the  geometry of the GRIFFIN HPGe crystals. 


\section{Analyses and Results}

The energy levels of $^{34}$Si were populated by the $\beta$ decay of the $4^-$ ground state and the $1^+$ isomeric state of the parent nucleus, $^{34}$Al. The $1^+$ isomeric component of the beam was determined to be 14(4)$\%$.

\begin{figure*}
\begin{center}
\includegraphics[width=0.90\textwidth]{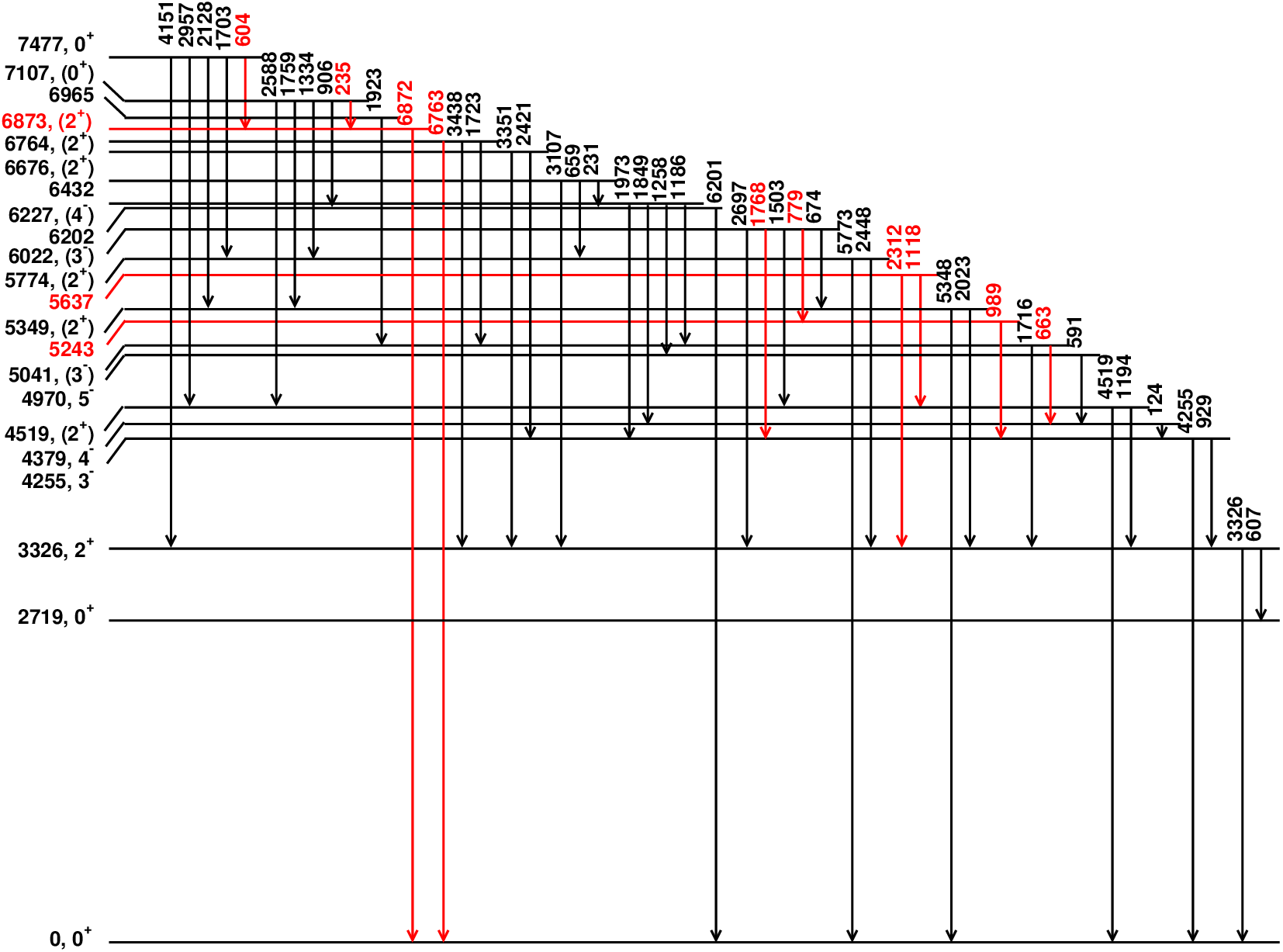}
\end{center}
\caption{Level scheme of $^{34}$Si built from the current analysis of the $\beta$-decay (both the ground state $4^-$ and the isomeric state $1^+$) of $^{34}$Al. The transitions and levels shown in red were observed for the first time in this work. The $\gamma$-ray branching ratios from the individual levels are listed in Table \ref{tab:exptab1}.}
\label{fig:lvlsch}
\end{figure*}

\begin{figure}
\begin{center}
\includegraphics[width=1.0\textwidth]{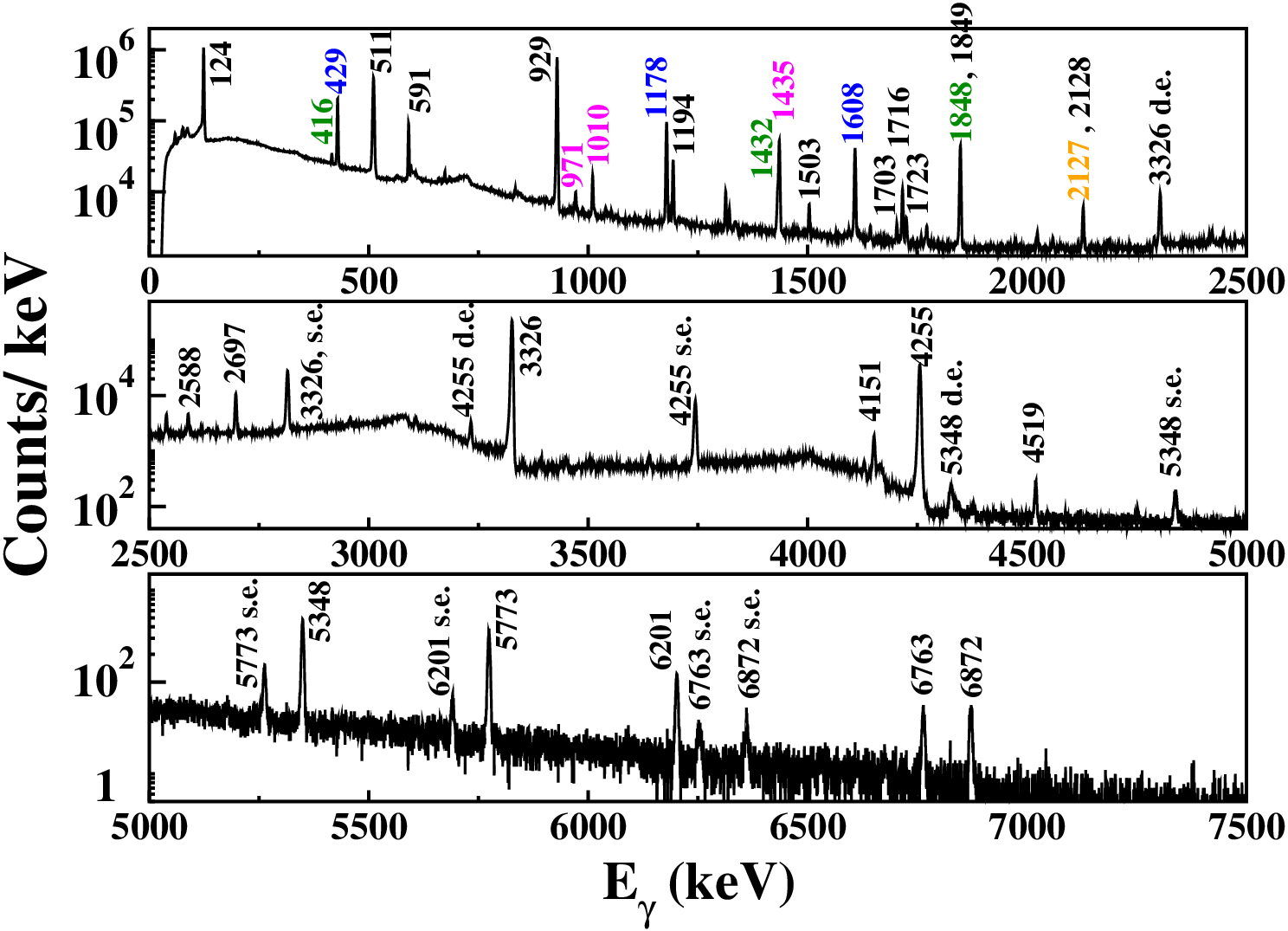}
\end{center}
\caption{$\beta$-gated $\gamma$-ray energy spectrum in coincidence with the SCEPTAR and ZDS detectors. Peaks belong to (a) $^{34}$Si are labeled in black (b) $^{33}$Si are labeled in magenta (c) $^{34}$P are labeled in blue (d) $^{33}$P are labeled in green and (e) $^{34}$S are labeled in orange. Peaks labeled with s.e. are the single-escape and those labeled with d.e. are the double-escape peaks.}
\label{fig:betasingles}
\end{figure}

The level scheme generated from the current analysis is presented in Fig. \ref{fig:lvlsch}. All the $\gamma$-ray transitions and the energy levels that were reported in Ref. \cite{lica34Si} have been observed in the current analysis. Ten newly identified $\gamma$-ray transitions and three newly observed energy states have been added to the level scheme which are shown in red. The  $\gamma$-ray spectrum of up-to 7.5 MeV energy in coincidence with the $\beta$ particles is shown in Fig. \ref{fig:betasingles}. 

\begin{figure}
\begin{center}
\includegraphics[width=1.0\textwidth]{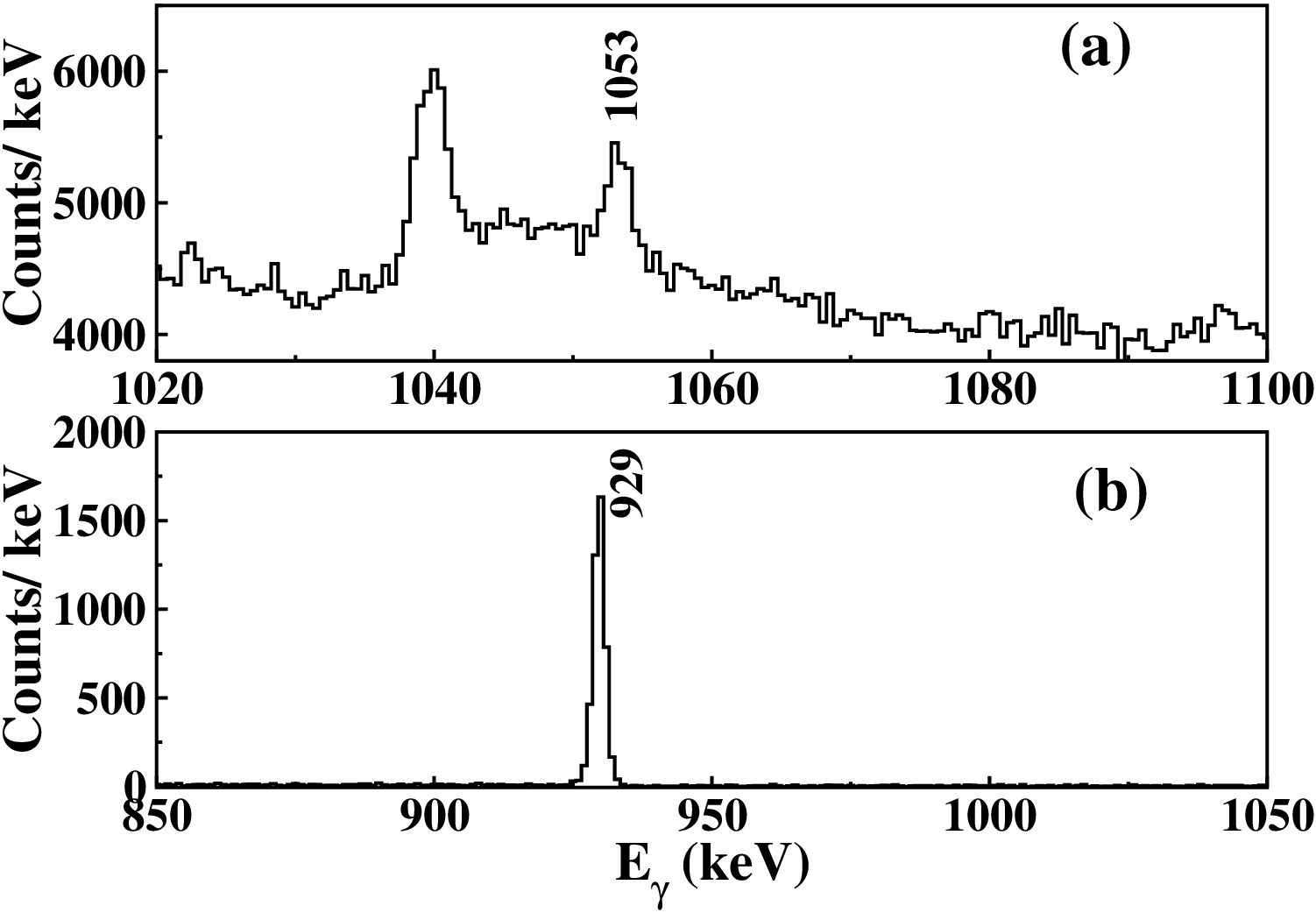}
\end{center}
\caption{(a) 1053 keV peak in the singles $\beta$-gated $\gamma$-ray spectrum (b) projection of the $\gamma$-$\gamma$ 180 degree angle matrix gated on the 124-keV transition showing the 929 keV peak, which is in immediate coincidence with the gating transition. See Ref. \cite{griffin} for details about summing correction method.}
\label{fig:sum1053}
\end{figure}

A transition at 1053 keV was reported first by Nummela $et \, al.$ \cite{nummela34Si} to be emitted from the level at 4379 keV along with a second $\gamma$-ray transition at 124 keV. This state first observed in that study was suggested to have spin-parity $3^-$, $4^-$ or $5^-$. The level was later observed along with the two transitions by Han $et \,al.$ \cite{han34Si} in a $\beta$-decay study. In a most recent study \cite{lica34Si} both the $\gamma$-ray transitions from this level were reported and the 1053 keV transition was reported to result from the summing of the 124 and 929 keV transitions. However, no justification was presented to support this argument. The level was assigned spin-parity $(4^-)$ by that work.  Fig. \ref{fig:sum1053} shows the HPGe spectra used to determine the summing-in correction.  After the summing-in and summing-out corrections for the 1053 keV peak, the total number of counts obtained were 5(127). This confirms that the 1053 keV peak is a summed peak from the contributions of 124 and 929 keV $\gamma$-ray transitions. This result is of significant importance in the assignment of the spin and parity of the 4379 keV level which is discussed later.

\begin{figure}
\begin{center}
\includegraphics[width=1.0\textwidth]{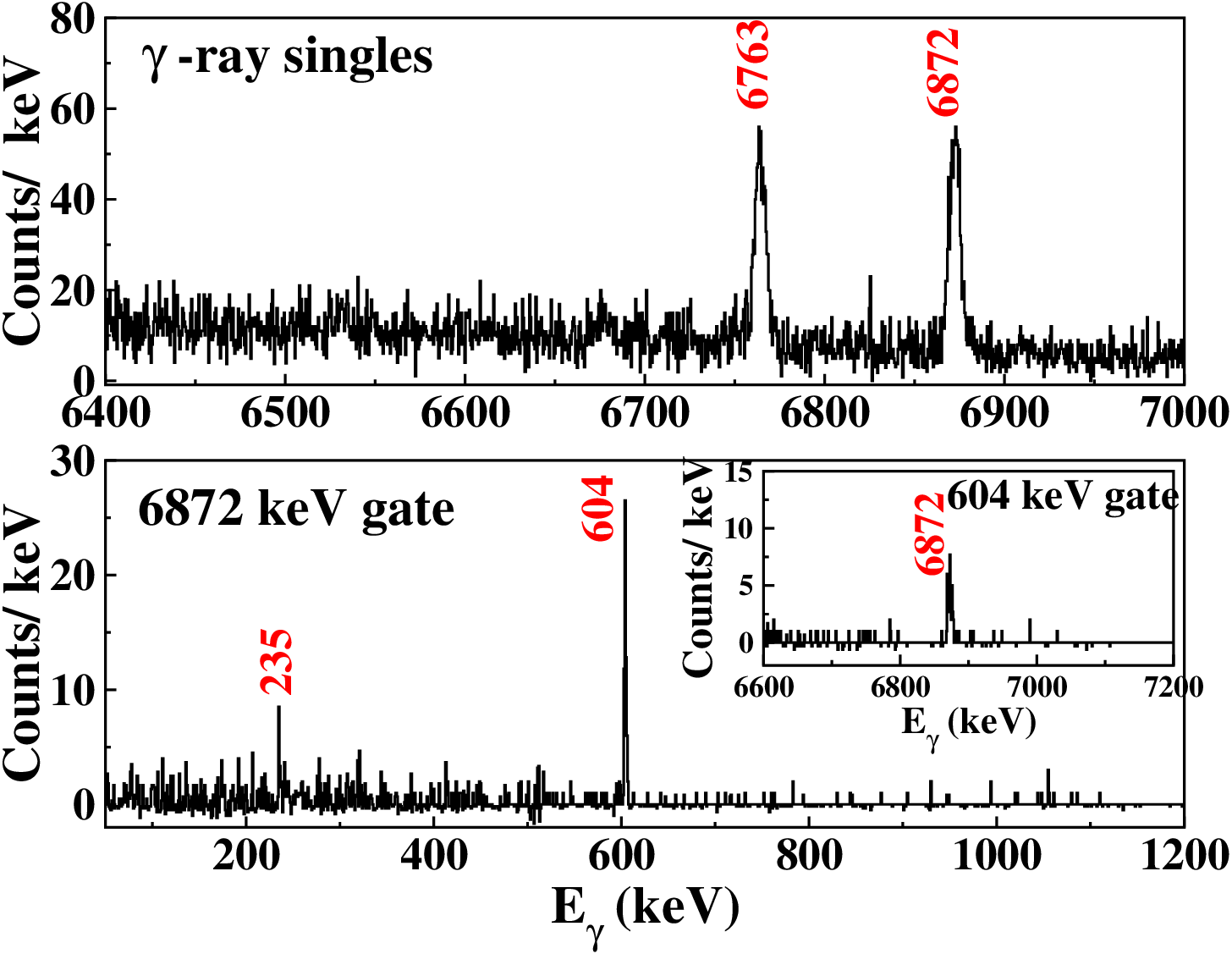}
\end{center}
\caption{Two newly observed ground state $\gamma$-ray transitions in $^{34}$Si and the corresponding coincidence. The red text indicates that the $\gamma$-ray peaks were newly identified.}
\label{fig:newP1}
\end{figure}

In the present work, $\gamma$-ray  transitions up to about 7 MeV were observed. Fig. \ref{fig:newP1} shows two such ground state transitions at 6763 and 6872 keV. The transition at 6763 keV decays from a previously reported state at 6764 keV, whereas the 6872 keV $\gamma$-ray decays from a newly identified level at 6872 keV. The $\gamma$-ray at 6872 keV is in coincidence with two newly observed transitions at 604 and 235 keV which decay from the previously known 7107 and 7477 keV levels respectively, as shown in Fig. \ref{fig:newP1}. The ground state transitions at 6763 and 6872 keV suggest the corresponding levels have low spins and are populated by the $\beta$-decay of the $1^+$ isomer. Therefore, spin-parity $2^+$ describe the levels best, though the possibility of $1^+$ can not be ruled out. Two very weak transitions at 1118 and 2312 keV were observed for the first time in the current work leading to a new state at 5637 keV. The associated $\gamma$-ray peaks were seen only in the coincident spectra gated on the 1194 and 3326 keV transitions, respectively. The $\gamma$-ray transitions at 1849 and 2128 keV, which were previously observed and also confirmed in the current analysis, are doublets with the transitions that belong to $^{33}$P and $^{34}$S respectively.

Table \ref{tab:exptab1} lists the energy levels, $\gamma$-ray intensities relative to the 3326 keV transition, $\gamma$-ray transition branching ratios from the individual levels and spin-parities deduced and suggested from the current experimental analysis. The branching ratios of the $\gamma$-ray transitions have been compared to those extracted from Ref. \cite{lica34Si} in this table. In most cases, the ratios are consistent between the two works. A large discrepancy in the branching ratio from the previous measurement is observed in the 6764 keV state. The current analysis reports a ground state transition at 6763 keV, which is the strongest branch for this level, whereas this transition was not observed in the previous work. The summing correction has confirmed this $\gamma$-ray is real. 


\begin{table*}
\setlength{\tabcolsep}{0.7em}
\caption{Observed excitation energies ($\text E_x$) and spins ($\text J^\pi$) together with the associated $\beta$-delayed $\gamma$-ray transitions ($\text E_\gamma$) of $^{34}$Si deduced from the current experimental analysis are presented in this table. The newly observed states and transitions are shown in boldface. The intensities ($\text I_\gamma$) of the $\gamma$-rays relative to that of 3326 keV and the branching ratios ($\text I_\gamma$ (BR)) from the individual levels are also shown along with the comparison to the literature values. All the $\gamma$ ray intensities are corrected for the summing effect and the states are corrected for the $\gamma$-ray recoil energies.}
\label{tab:exptab1} 
\begin{tabular}{|c|c|c|c|c|c|c|c|}
\hline 
{{$\text E_x$ (keV)}} & {{$\text J^\pi _i$}} & {{$\text E_\gamma$ (keV)}} & {Relative intensity $\text I_\gamma$} & {{$\text J^\pi _f$}} & {{$\text I_\gamma$ (BR)}} & {{$\text I_\gamma$ (BR)\cite{lica34Si}}} \\[0.5ex]  \hline  
3325.7(2) & $2^+$   & 3325.5(2)     & 100.0(15)   & $0^+$                     & 100.0(15) & 100 \\ 
          &         & 607.0(3)          & 0.15(2)  & $0^+$                     & 0.15(2)  & 0.06      \\ \hline
4254.8(2) & $3^-$   & 929.2(2)      & 89.9(14)    & $2^+$                     & 100.0(15) & 100       \\ 
          &            & 4254.6(2)      & 23.8(4)   & $0^+$                     & 26.4(4) & 29.8         \\ \hline
4378.7(2) & $4^-$      & 124.0(2)       & 42.6(6)    & $3^-$                     & 100   & 100      \\ \hline
4519.2(2) & $(2^+)$      & 1193.5(2)      & 3.36(5)   & $2^+$                     & 100.0(15) & 100      \\
          &            & 4518.8(4)      & 0.15(1)   & $0^+$                     & 4.4(2)   & 6.0    \\ \hline
4969.6(3) & $5^-$      & 591.0(2)       & 6.4(1)    & $4^-$                     & 100   & 100    \\ \hline
5041.4(2) & $(3^-)$    & \textbf{662.7(2)}      & 0.09(1)    & $4^-$            & 5.7(7)   &     \\ 
          &            & 1715.8(2)      & 1.68(3)   & $2^+$                     & 100.0(17) & 100       \\ \hline
\textbf{5243.3(3)} &            & \textbf{988.5(2)}     & 0.10(1)   & $3^-$    & 100   &     \\ \hline
5348.8(2) & $(2^+)$    & 2023.2(3)      & 0.22(1)   & $2^+$                     & 50(2) & 47       \\ 
          &            & 5348.3(3)      & 0.45(1)   & $0^+$                     & 100(2) & 100       \\ \hline
\textbf{5637.4(5)} &            & \textbf{1117.9(3)}       & 0.03(1) & $(2^+)$   & 52(13)   &     \\
		  &            & \textbf{2312.3(5)}       & 0.06(2)  & $2^+$    & 100(41) &     \\ \hline
5773.5(2) & $(2^+)$    & 2447.7(2)      & 0.19(2)   & $2^+$                     & 51(5)    & 55     \\ 
          &            & 5773.0(3)      & 0.38(1)   & $0^+$                     & 100.0(23)  & 100         \\ \hline
6022.4(2) & $(3^-)$    & 673.5(2)       & 0.49(1)    & $(2^+)$                   & 17.0(4) & 16.6       \\ 
          &            & \textbf{778.9(2)}      & 0.13(1)    &                  & 4.7(3)  &      \\ 
          &            & 1503.2(2)      & 0.73(1)   & $(2^+)$                   & 25.6(5) & 29.9      \\ 
          &            & \textbf{1767.6(3)}     & 0.12(4)   & $3^-$            & 4.4(15)    &    \\ 
          &            & 2696.5(2)      & 2.86(5)   & $2^+$                     & 100.0(16) & 100        \\ \hline
6201.7(4) &            & 6201.1(4)      & 0.15(1)   &    $0^+$                       & 100 & 100       \\ \hline
6227.4(2) & $(4^-)$    & 1185.9(2)      & 0.18(1)   & $(3^-)$                   & 35(2) & 29       \\ 
          &            & 1257.9(2)      & 0.06(1)   & $5^-$                     & 12(2)  & 5      \\
          &            & 1848.7(2)      & 0.52(2)   & $4^-$                     & 100(4)   & 100    \\ 
          &            & 1972.5(2)      & 0.09(1)   & $3^-$                     & 17(2)  & 16      \\ \hline
6432.4(2) &            & 230.7(3)       & 0.03(1)    &                           & 7(2)  & 10     \\ 
          &            & 658.8(2)       & 0.18(1)    & $(2^+)$                   & 38(3)   & 31     \\
          &            & 3106.6(2)      & 0.46(2)   & $2^+$                     & 100(4)   & 100    \\ \hline
6676.3(2) & $(2^+)$    & 2421.3(2)      & 0.23(1)   & $3^-$                     & 100(4)   & 100     \\
          &            & 3350.6(2)      & 0.10(1)   & $2^+$                     & 45(2) & 39        \\ \hline
6764.0(2) & $(2^+)$    & 1722.5(3)      & 0.04(1)   & $(3^-)$                & 45(9)   & 100     \\ 
          &            & 3438.2(3)      & 0.05(1)   & $2^+$                  & 59(9)   & 93     \\ 
          &            & \textbf{6763.3(4)}     & 0.08(1)   & $0^+$         & 100(5)   &     \\ \hline
\textbf{6872.5(4)} & $(2^+)$ & \textbf{6871.8(4)}       & 0.09(1)   & $0^+$         & 100 &       \\ \hline
6964.6(6) &            & 1923.1(6)      & 0.03(1)   & $(3^-)$                & 100   & 100     \\ \hline
7107.4(2) & $(0^+)$      & \textbf{235.1(3)}        & 0.011(3)    &               & 1.3(4)  &       \\ 
          &            & 905.5(2)       & 0.10(1)    &                & 12(2)    & 13   \\
          &            & 1333.9(8)      & 0.06(1)   & $(2^+)$                & 8(2)    & 12    \\ 
          &            & 1758.6(2)      & 0.13(1)   & $(2^+)$                & 16(1) & 12       \\ 
          &            & 2588.0(2)      & 0.82(2)   & $(2^+)$                & 100(2) & 100      \\  \hline
7476.5(2) & $0^+$      & \textbf{603.8(2)}      & 0.07(1)    &               & 7(1)   &     \\
          &            & 1703.0(2)      & 0.27(1)   & $(2^+)$                & 29(1)  & 29      \\ 
          &            & 2127.8(3)      & 0.07(2)   & $(2^+)$                & 8(2)     & 7   \\ 
          &            & 2957.2(2)      & 0.40(2)   & $(2^+)$                & 42(2)    & 38    \\ 
          &            & 4150.5(2)      & 0.95(4)   & $2^+$                  & 100(4)    & 100    \\ \hline
\end{tabular}
\end{table*}


Since the $\beta$-decay  of the $4^-$ ground and the $1^+$ isomeric states of $^{34}$Al directly populate different levels in $^{34}$Si, it  is possible to determine the $\beta$-feeding intensities for both decays from the relative $\gamma$-ray intensities listed in Table \ref{tab:exptab1}. The $\beta$-delayed one  neutron emission probabilities, $P_n$,  for the $4^-$ ground and the $1^+$ isomeric states were determined previously  \cite{lica34Si} to be 22(5)$\%$ and 11(4)$\%$ respectively.  As a result, the sum of the relative $\beta$-feeding intensities for the negative parity states in $^{34}$Si populated in the $\beta$-decay of the $^{34}$Al $4^-$ ground state was normalized to 78$\%$. The direct $\beta$-feeding from the $^{34}$Al isomeric $1^+$ level to the $0^+$ ground and first excited $0^+$ states in $^{34}$Si could not be determined in the present study. Since direct $\beta$-feeding of the $0^+_1$ and $0^+_2$ levels in $^{34}$Si by the $1^+$ isomeric state in $^{34}$Al were reported to be 27(10)$\%$ and 37(6)$\%$, respectively \cite{lica34Si}, the normalization factor for this decay to positive parity levels in $^{34}$Si observed in the present experiment should be 25$\%$.  However, since the $1^+$ isomeric content of the beam was only 14(4) $\%$ and ~80-90$\%$ of the gamma-ray intensity in both $\beta$-decays feeds through the 3326 keV $2^+$ in $^{34}$Si, the uncertainty in the direct $\beta$-feeding intensity of this level was very large.  As a result, we chose to normalize the sum of the relative $\beta$-feeding intensities for the positive parity states in $^{34}$Si such that the value obtained for the 7477 keV $0^+$ level agreed with the one determined previously \cite{lica34Si}. The resulting $\beta$-feeding intensities and experimental log$ft$ values for both decays are listed in Table II together with the results reported previously \cite{lica34Si}.  In general the present results are in good agreement but more precise than the previous data for the $4^-$ ground state decay.

\begin{table*}
\setlength{\tabcolsep}{0.7em}
\caption{Observed excitation energies, $\beta$ feeding intensities ($\text I_\beta$) and the corresponding log$ft$ values are tabulated below and compared with those reported previously \cite{lica34Si}. In the present work the uncertainties in the $\beta$ feeding intensities only include those associated with the measured relative $\gamma$-ray intensities. The $^{34}$Al $4^-$ ground state Q value, 16.957(14) MeV \cite{qvalue}, and the $1^+$ isomeric state excitation energy, 46.7 keV \cite{lica34Al}, together with $\beta$-decay half-lives of 53.73(13) ms and 22.1(2) ms for the $4^-$ ground state and $1^+$ isomeric state in $^{34}$Al \cite{lica34Si}, respectively, were used to determine the log$ft$ values. }
\label{tab:exptab2}
\begin{tabular}{|c|c|c|c|c|c|c|c|c|}
\hline
\multirow{2}{*}{$\text E_x$ (keV)}        & \multicolumn{4}{c|}{Populated by $4^-$}            & \multicolumn{4}{c|}{Populated by $1^+$}           \\ \cline{2-9}
                                     & $\text I_\beta \%$ & log$ft$ & $\text I_\beta \%$ \cite{lica34Si} & log$ft$ \cite{lica34Si} & $\text I_\beta \%$  & log$ft$ & $\text I_\beta \%$ \cite{lica34Si} & log$ft$ \cite{lica34Si}\\ \hline
3325.7(2)                          	&           		 	&         				& $<$ 0.1       	&             			& 1.4(64) 		 		&  $>$5.34       	& 7(3)          		& 5.4(2)      \\ \hline
4254.8(2)                        	 	& 46.1(12)  	&  4.79(1)       	& 41(7)         		& 4.84(8)     		&            					&         			   	& $<$0.1      	&             \\ \hline
4378.7(2)	                        	& 23.3(5)    	&  5.06(1)        	& 28(3)         		& 4.98(5)     		&            					&         				&               			&             \\ \hline
4519.2(2)                            	&           	    	&         				& $<$0.1      		&             			& 5.0(2)     			&  5.33(2)       	& 4.8(12)       	& 5.4(1)      \\ \hline
4969.6(3)                            	& 4.2(1)      	&  5.71(1)       	& 3.9(4)        		& 5.74(5)     		&            					&         				&               			&             \\ \hline
5041.4(2)                            	& 1.00(3)   	&   6.32(1)      	& 0.95(13)      	& 6.34(6)     		&            					&         				& $<$0.1      		&             \\ \hline
\textbf{5243.3(3)}  		  	& $<$ 0.01   &         				    &               			&             			&            					&         				&               			&             \\ \hline
5348.8(2)                            	&           			&         				&               			&             			&  $<$0.1		        &         				& $<$0.1      		&             \\ \hline
\textbf{5637.4(5)} 		  	&           			&         				&               			&             			& 0.3(1)     			&  6.4(2)       	&               			&             \\ \hline
5773.5(2)                           	&           			&         				&              		 	&             			& 0.2(1)     			&  6.5(2)       	& $<$0.4      		& $>$6.8    \\ \hline
6022.4(2)	                         	& 2.84(6)   	&   5.69(1)      	& 2.76(21)      	& 5.70(4)     		&            					&         				&               			&             \\ \hline
6201.7(4)                           	&           			&         				&               			&             			& $<$0.1    		    &   	      				& $<$0.1      		&             \\ \hline
6227.4(2)	                         	& 0.56(2)   	&   6.36(2)      	& 0.79(7)       	& 6.21(4)     		&            					&         				&               			&             \\ \hline
6432.4(2)  	                       	&           			&         				&               			&             			& 2.1(1)     			&  5.37(2)       	& 2.1(2)        		& 5.36(5)     \\ \hline
6676.3(2)                            	&          	 		&         				&               			&             			& 1.07(3)    			& 5.62(1)        	& 1.0(1)        		& 5.64(5)     \\ \hline
6764.0(2)	                         	&           			&         				&               			&             			& 0.52(3)    			&  5.92(3)       	& 0.26(4)       	& 6.2(1)      \\ \hline
\textbf{6872.5(4)}  		  	&      				&         				&               			&             			& $<$0.1    		    &   				      	&               			&             \\ \hline
6964.6(6)                           	&           			&         				&               			&             			& 0.09(2)    			&  6.63(10)       & $<$0.1      		&             \\ \hline
7107.4(2)	                        	&          	 		&         				&               			&             			& 3.6(1)    		 		&  4.99(1)       	& 3.4(4)        		& 5.0(1)      \\ \hline
7476.5(2)	                        	&           			&         				&               			&             			& 5.6(1)     			&   4.72(1)      	& 5.6(5)        		& 4.72(4)  \\
\hline
\end{tabular}
\end{table*}

\begin{table*}
\setlength{\tabcolsep}{0.7em}
\caption{Summary of the results from $\gamma$-$\gamma$ angular correlation analysis. The mixing $\delta 2$ for the  $J_m \rightarrow J_f$ (middle to final) transition was kept fixed in order to calculate the mixing $\delta 1$ for the $J_i\rightarrow J_m$ (initial to middle) transition in each cascade.}
\label{tab:exptab3}
\begin{tabular}{|c|c|c|c|c|c|}
\hline 
{$\text E_x$ (keV)}                   & {$J_i\rightarrow J_m \rightarrow J_f$} & $\text E_{\gamma 2}$ (keV)       & $\text E_{\gamma1}$ (keV)        & $\delta_{\gamma 2}$  & $\delta_{\gamma 1}$ \\[0.5ex] \hline 
4255                  							& $3^- \rightarrow 2^+ \rightarrow 0^+$ & 3326                  & 929       		& $0$          & $0.02(1)$     \\ \hline
4379                					 		 	& $4^- \rightarrow 3^- \rightarrow 2^+$ & 929                  	& 124            	& $0.02(1)$      & $-0.012(17)$  \\ \hline
4519                  							& $(2^+) \rightarrow 2^+ \rightarrow 0^+$ & 3326                & 1194      		& $0$           & $0.43(2)$   \\ \hline
\multirow{2}{*}{4970} 				& $5^- \rightarrow 4^- \rightarrow 3^-$ & \multirow{2}{*}{124} & \multirow{2}{*}{591} & $-0.012(17)$ & $0.01(4)$  \\ \cline{2-2} \cline{5-5} 
                      									& $3^- \rightarrow 4^- \rightarrow 3^-$ &           &                 & $-0.012(17)$     & $-0.09(35)$ \\ \hline
\end{tabular}
\end{table*}

\begin{figure}
\begin{center}
\includegraphics[width=0.8\textwidth]{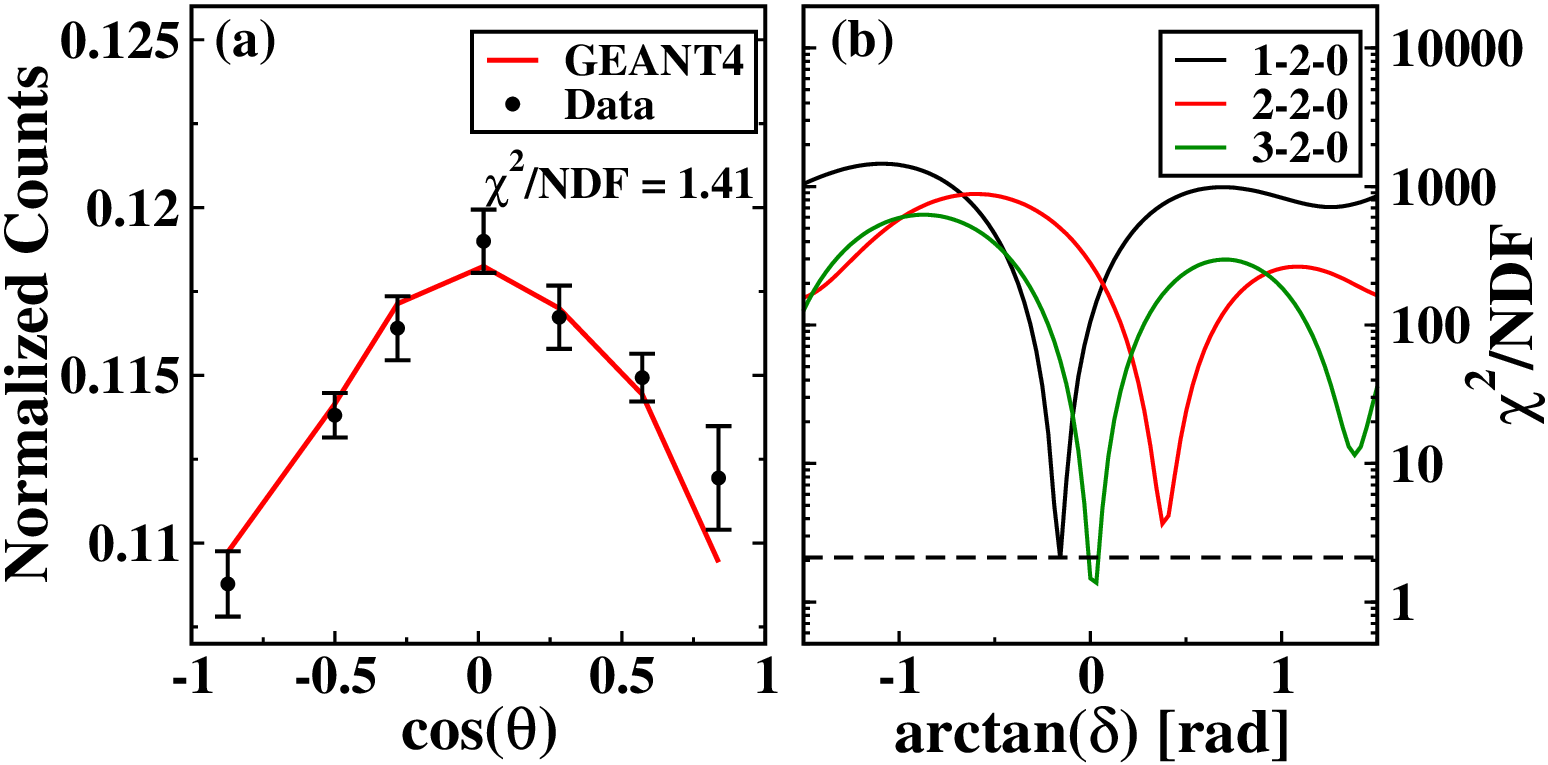}
\end{center}
\caption{(a) Measured $\gamma$-$\gamma$ angular correlation for the 929-3326 keV transitions associated with the 4255-3326-0 keV cascade. The linear combination of the simulated correlation performed with GEANT4 is shown in red.(b)  $\chi^2$ fit to the measured $\gamma$-$\gamma$ angular correlation for different spin hypotheses. The mixing ratio of the 3326 keV transition was kept fixed at $0$ and $\delta$ along the x axis in this plot is the mixing of the 929 keV $\gamma$-ray transition. The value of $\delta$ is extracted as $0.02(1)$. The $\chi^2$ analysis confirms the well established spin 3 to the 4255 keV level.  All the 51 angular bins have been grouped into seven as mentioned in the text.}
\label{fig:ac4255}
\end{figure}

\begin{figure}
\begin{center}
\includegraphics[width=0.8\textwidth]{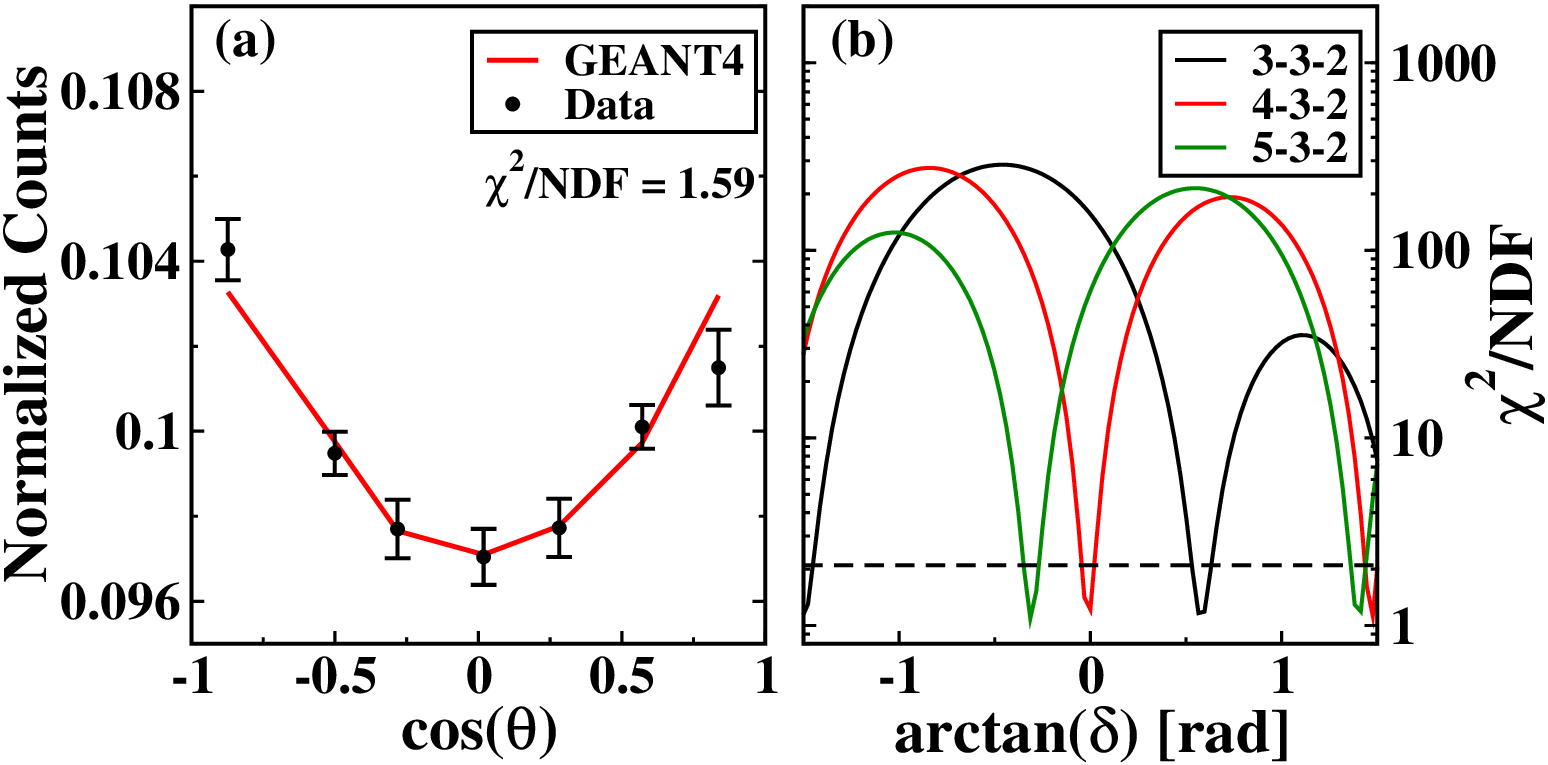}
\end{center}
\caption{(a) Measured $\gamma$-$\gamma$ angular correlation for the 124-929 keV transitions associated with the 4379-4255-3326 keV cascade. The linear combination of the simulated correlation performed with GEANT4 is shown in red. (b) $\chi^2$ fit to the measured $\gamma$-$\gamma$ angular correlation for different spin hypotheses. In this plot the mixing ratio of the 929 keV transition was kept fixed at $0.02$ and $\delta$ is the mixing of the 124 keV $\gamma$-ray transition. The $\delta_{\gamma 1}$ value for the 124 keV transition is extracted as 0.012(17) for the spin hypothesis of 4.  The uncertainty in $\delta_{\gamma 1}$ includes the uncertainty in $\delta_{\gamma 2}$ determined for the 929 keV transition. The spin assignment for the corresponding level at 4379 keV is discussed in the main text.}
\label{fig:ac4379}
\end{figure}

One of the main goals of the current work was to conduct the $\gamma$-$\gamma$ angular correlation analysis in order to make spin assignments to the energy levels of $^{34}$Si. The results of this analysis are summarized in Table  \ref{tab:exptab3}. The level at 4255 keV was first observed by Baumann $et \, al.$ \cite{baumann34Si} via the $\beta$-decay of $^{34}$Al and a $3^-$ spin-parity was assigned based on the  $\beta$-feeding intensity. In the present work, the $\gamma$-$\gamma$ angular correlation between the 929 and 3326 keV transitions associated with the 4255-3326-0 keV energy-level cascade was determined. Fig. \ref{fig:ac4255} shows the measured $\gamma$-$\gamma$ angular correlation fitted with the GEANT4 simulation and the $\chi ^2$ plot for different spin hypotheses for the 4255 keV level. The dotted line represents a 95$\%$ confidence limit. From this figure, it is clear that 3 is the only favored spin for the state. Though a $3^-$ spin-parity is very obvious from the $\beta$-feeding intensities reported previously and the feeding-decay patterns of the $\gamma$-ray transitions to-and-from this level, here we confirm the spin with the $\gamma$-$\gamma$ angular correlation analysis. The M2/E1 mixing ratio was calculated as $0.02(1)$ for the 929 keV transition. 

\begin{figure}
\begin{center}
\includegraphics[width=0.8\textwidth]{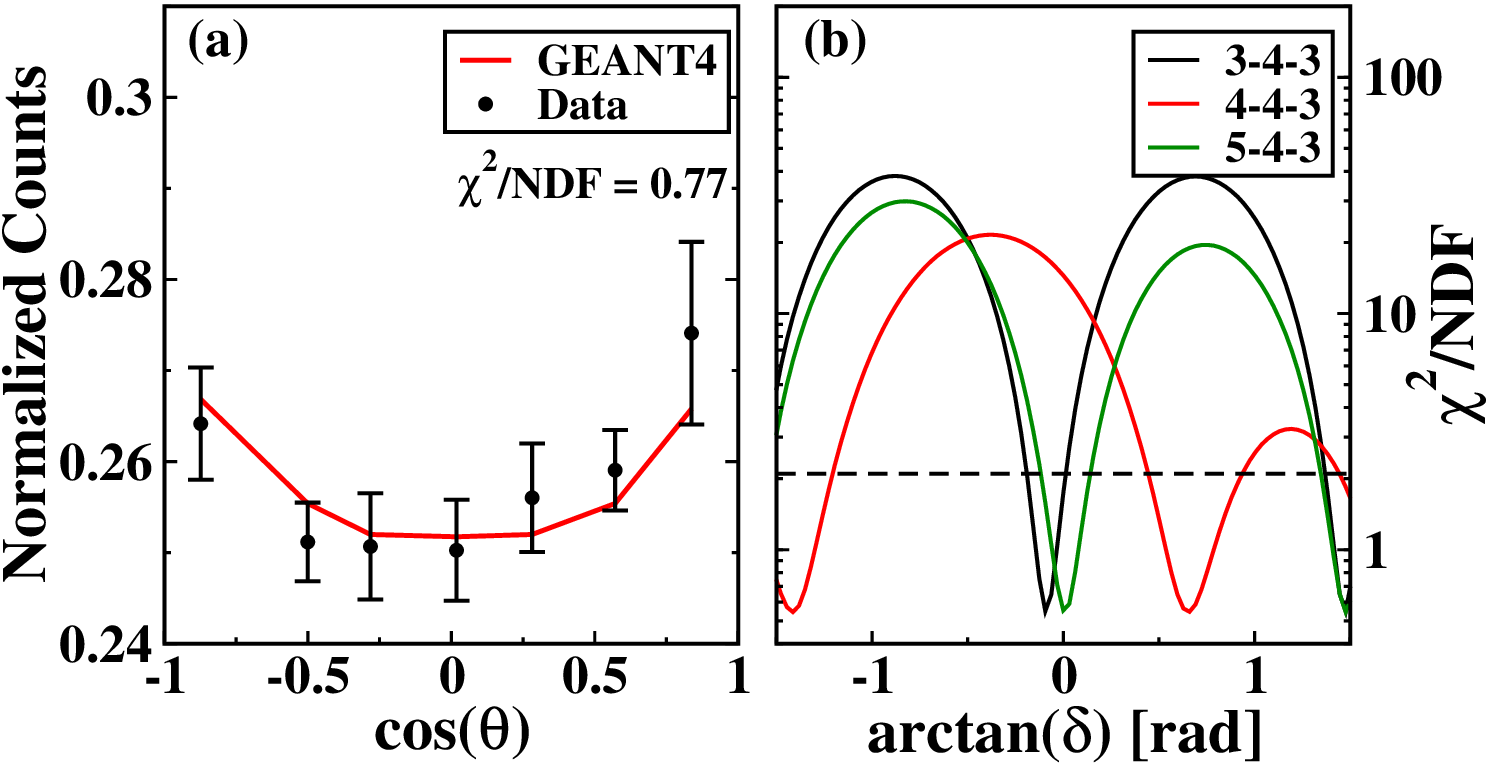}
\end{center}
\caption{(a) Measured $\gamma$-$\gamma$ angular correlation for the 591-124 keV transitions associated with the 4970-4379-4255 keV cascade. The linear combination of the simulated correlation performed with GEANT4 is shown in red. (b) $\chi^2$ fit to the measured $\gamma$-$\gamma$ angular correlation for different spin hypotheses. In this plot the mixing ratio of the 124 keV transition was kept fixed at $-0.012$ and $\delta$ is the mixing of the 591 keV $\gamma$-ray transition.  The uncertainty in $\delta_{\gamma 1}$ includes the uncertainty in $\delta_{\gamma 2}$ determined for the 124 keV transition.  The value of $\delta$ is extracted as $0.01(4)$ for the spin hypothesis 5 which is consistent with a pure M1 type and $-0.09(35)$ for the spin hypothesis of 3 to the 4970 keV level.}
\label{fig:ac4970}
\end{figure}

\begin{figure}
\begin{center}
\includegraphics[width=0.8\textwidth]{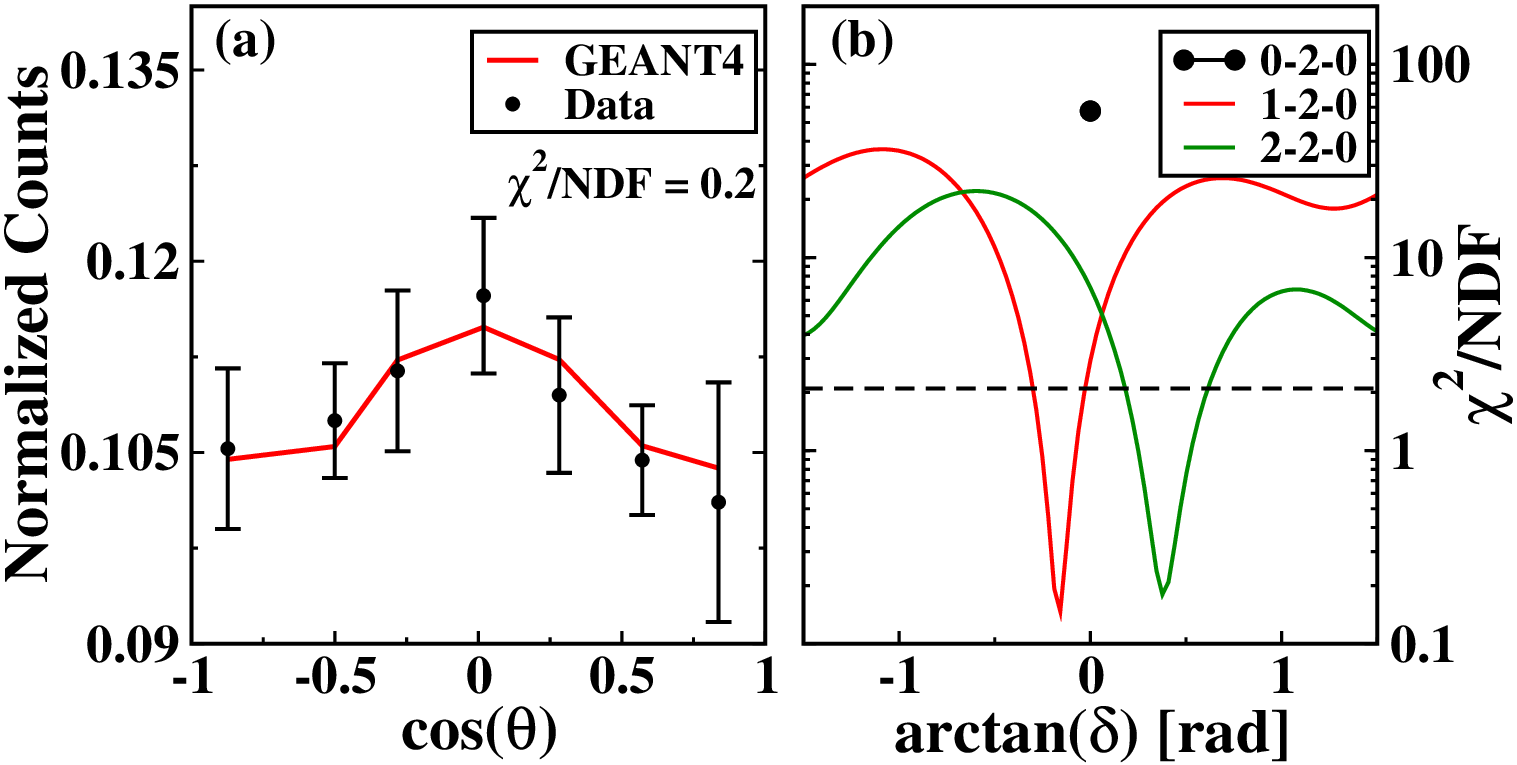}
\end{center}
\caption{(a) Measured $\gamma$-$\gamma$ angular correlation for the 1194-3326 keV transitions associated with the 4519-3326-0 keV cascade. The linear combination of the simulated correlation performed with GEANT4 is shown in red. (b) $\chi^2$ fit to the measured $\gamma$-$\gamma$ angular correlation for different spin hypotheses. The mixing ratio of the 3326 keV transition was kept fixed at $0$ and $\delta$ in this plot is the mixing of the 1194 keV $\gamma$-ray transition. The value of $\delta$ is extracted as $0.43(2)$ for the spin hypothesis of $2$ to the 4519 keV level. Discussions are in the main text.}
\label{fig:ac4519}
\end{figure}

The 4379 keV state was reported to have a negative parity in the previous $\beta$-decay studies \cite{baumann34Si, timis34Si, han34Si, lica34Si}. Among them, Nummela $et \, al.$ and Han $et\,  al.$ have reported the $\gamma$-ray transition at 1053 keV and suggested $(3^-, \, 4^-, \, 5^-)$ spin-parity as previously mentioned. Timis $et\, al.$ have also suggested either of $(3^-, \, 4^-, \, 5^-)$ spin-parity. In the latest work done by Lic\u{a} $et \, al.$, $(4^-)$ was tentatively assigned. In the current work, we have performed $\gamma$-$\gamma$ angular correlation between the 124 and 929 keV transitions as shown in Fig. \ref{fig:ac4379}(a). From the $\chi^2$ plot presented in \ref{fig:ac4379}(b), it can be seen that all the spin hypotheses in the picture are possible. However, for $J_i=3$, the  E2/M1 mixing ratio is large, which is less likely for a very low energy $\gamma$-ray at 124 keV. Moreover, according to the Weisskopf single particle estimation, the transition rates between the 124 and 1053-keV peaks would be comparable if 4379 keV level were a $3^-$, which is clearly not the case as we have already confirmed that 1053 keV transition is a summed peak. 
On the other hand, almost no mixing is expected for an E2 type transition, which would be the case if the energy level had $5^-$ spin-parity, which is not satisfied in the angular correlation analysis. Hence, spin 4 with a negative parity is the only possible choice consistent with very low mixing, calculated as $-0.012(17)$. 

\begin{figure}
\begin{center}
\includegraphics[width=0.8\textwidth]{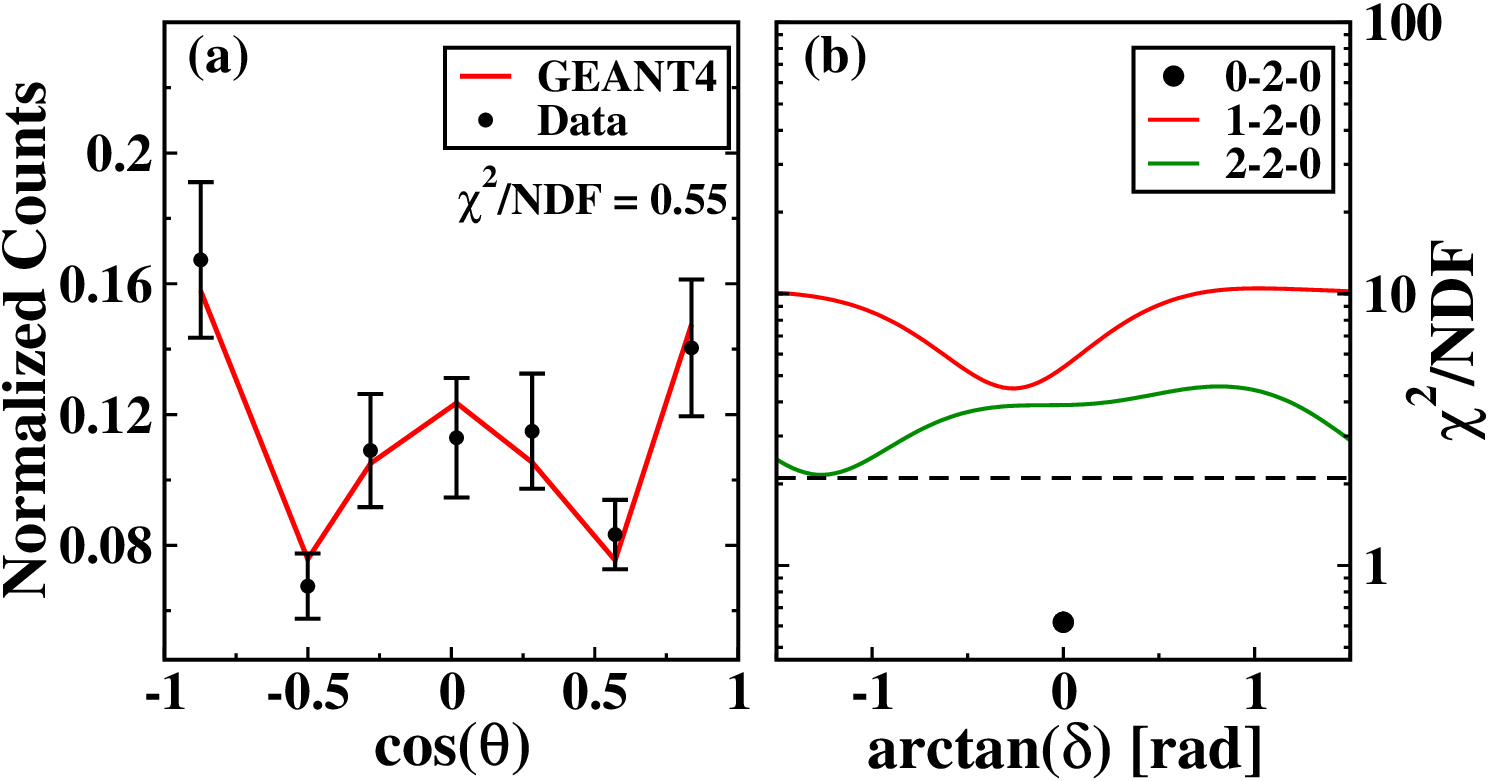}
\end{center}
\caption{(a) Measured $\gamma$-$\gamma$ angular correlation for the 4151-3326 keV transitions associated with the 7477-3326-0 keV cascade. The linear combination of the simulated correlation performed with GEANT4 is shown in red. (b) $\chi^2$ fit to the measured $\gamma$-$\gamma$ angular correlation for different spin hypotheses of the 7477 keV level. From this plot, it is clear that spin $0$ is the only possible option for this state which is in very good agreement with the theoretical predictions made in the current work.}
\label{fig:ac7477}
\end{figure}

The $\gamma$-$\gamma$ angular correlation has been performed between the 124 and 591-keV transitions to assign spin to the 4970 keV level as shown in Fig. \ref{fig:ac4970}(a), (b). This state was previously reported in the $\beta$-decay experiments of Refs. \cite{nummela34Si, han34Si, lica34Si} and was suggested to have $(3^-, \, 4^-, \, 5^-)$ spin-parity by Nummela $et \, al.$ and Han $et \, al.$, whereas, Lic\u{a} $et \, al.$ have called it a $(5^-)$. The $\chi^2$ plot in Fig. \ref{fig:ac4970}(b) shows that, based on the $\gamma$-$\gamma$ angular correlation analysis, a spin of 3, 4 or 5 is possible. A $4^-$ assignment is less likely because of the large E2/M1 mixing ratio. However, $3^-$ and $5^-$ can not be resolved from the experimental analysis. More discussion about this state is in the next section.

The level at 4519 keV was suggested to have $(2^+)$ spin-parity by Han and Lic\u{a} $et \, al.$ \cite{han34Si, lica34Si} based on the $\beta$-feeding intensities and the $\gamma$-ray transitions feeding to and decaying from this state. The $\gamma$-$\gamma$ angular correlation for the 1194-3326 keV cascade in this work shows that both spin 1 and 2 are possible, see Fig. \ref{fig:ac4519}. A $0^+$ spin-parity assignment is excluded by the $\gamma$-$\gamma$ angular correlation  for the 4519 keV level, consistent with the observed ground state transition.

The highest observed level at 7477 keV was assigned a tentative $(0^+)$ spin-parity in Ref. \cite{lica34Si} based on the argument that the state was populated directly by the decay of $1^+$ isomer of $^{34}$Al with a log$ft$ value of 4.72(4) and the pattern of the $\gamma$-rays decaying from it. We have measured the $\gamma$-$\gamma$ angular correlation for the 3326-4151 cascade, see Fig. \ref{fig:ac7477}(a), which  supports only the spin 0 hypothesis as can be seen in Fig. \ref{fig:ac7477}(b). Therefore, we confirm that the 7477 keV has a spin and parity $0^+$ whose dominant configuration will be discussed in the next section.

\section{Theoretical Discussion}

The experimental results on the structure of $^{34}$Si were compared with the predictions made by shell model calculations using the FSU interaction Hamiltonian \cite{lubna1, lubna2}. Shell-model code CoSMo \cite{cosmo} was utilized to perform these calculations. The FSU interaction covers a large part of the nuclear chart that includes nuclei ranging in mass number from around 10 to 50. It is a modern successor to a number of very successful effective interactions for individual shells supplemented with newly determined cross-shell matrix elements. 

\begin{figure*}
\begin{center}
\includegraphics[width=0.8\textwidth]{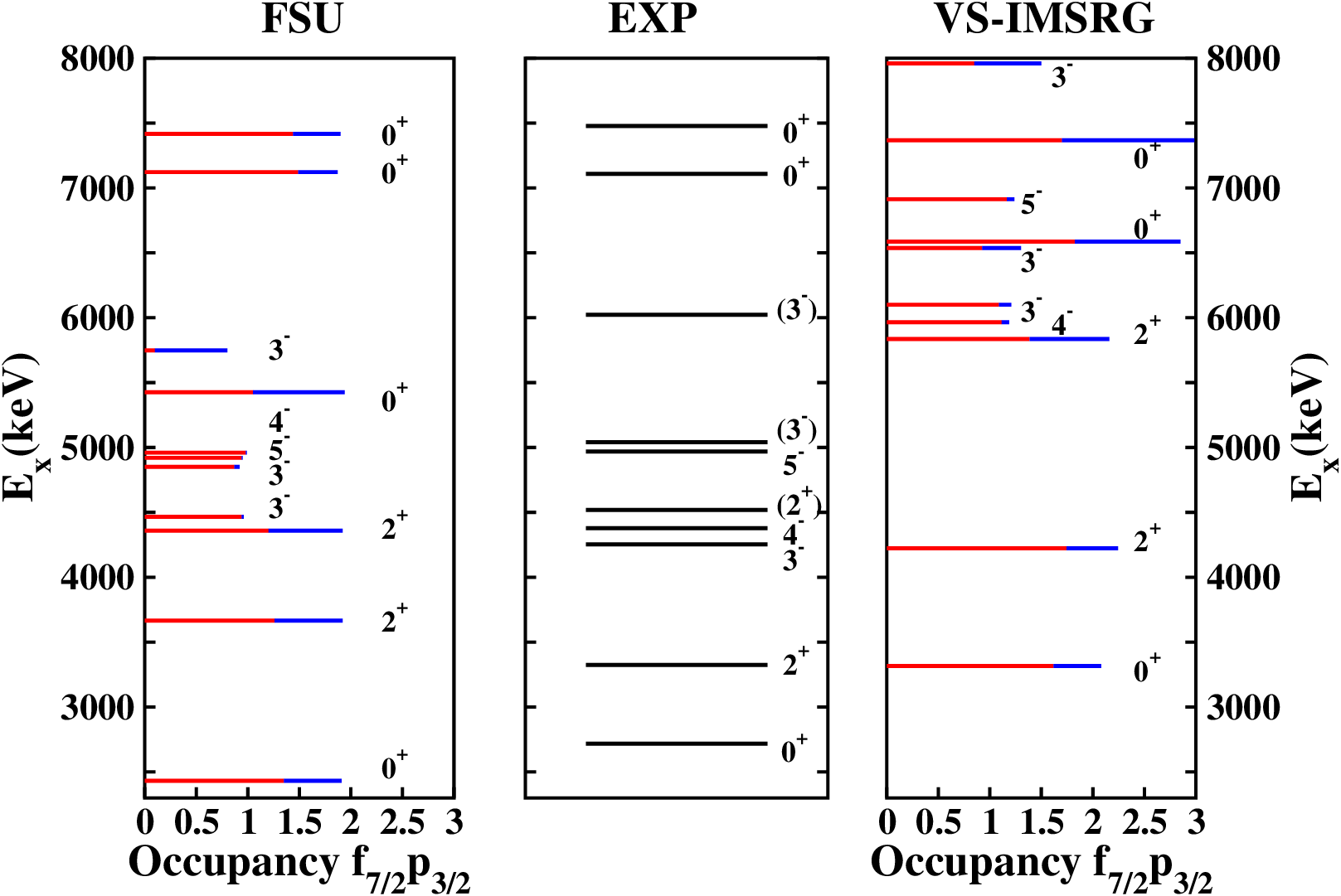}
\end{center}
\caption{Some observed excited energy states of $^{34}$Si are compared to those calculated with the FSU interaction and the VS-IMSRG method. The middle panel, which consists of the experimental states have the same energy scaling to that of the theory panels. In the calculated levels, the red portion represents the occupancy of the $0f_{7/2}$ orbital and the blue portion represents that of the $1p_{3/2}$ orbital.}
\label{fig:ThFsuAb}
\end{figure*}

Fig. \ref{fig:ThFsuAb} shows the comparison of some observed energy levels with the calculated states performed by using FSU interaction for different configurations with the $1\hbar\omega$ and $2\hbar\omega$ excitations. No mixing was considered in the calculations. All the positive parity energy states calculated with the FSU interaction shown in this figure are of $2\hbar\omega$ configuration within the $sdfp$ valence space, whereas the negative parity states are of $1\hbar \omega$ configuration which spans the $psdfp$ valence space.  The first excited $0^+$ and $2^+$ states were discussed to have the dominant $2\hbar\omega$ structure in Ref. \cite{rotaru}. The predictions made with the FSU interaction are in agreement with the arguments as shown in Ref. \cite{lubna2}. The suggested second experimental $(2^+)$ state at 4519 keV is most likely dominated by the $2\hbar\omega$ configuration, as it decays via the gamma transition at 1194 keV to the 3326 keV state of similar dominant configurations and by a weak transition of 4519 keV to the ground state. 

The observed experimental $0^+$ state at 7477 keV may correspond to one of the calculated levels at 7122 and 7614 keV, which are close in energies and are of $2\hbar \omega$ configurations. This is the highest observed experimental level and is populated by the $1^+$ isomeric state quite strongly with a log$ft$ value of 4.73(1). We have calculated the log$ft$ with the FSU interaction as listed in Table \ref{tab:thtab1}. The experimentally observed 7477 keV state has a better energy match to the predicted 7416 level but the log$ft$ value is closer to that predicted for the calculated energy at 7122 keV. Both the predicted levels have similar configurations. 

The positive parity states of $^{34}$Si were interpreted before with other configuration interactions as discussed in Refs. \cite{rotaru, lica34Si, Poves34Si,  nummela34Si}. Among them, SDPF-U-MIX quite successfully explained the positive-parity states, especially the lowest-lying $0^+$ and $2^+$ levels which are dominated by the intruder configurations and the B(E2) values, considering mixing between 0p0h and 2p2h excitations \cite{rotaru}. Direct comparison of different theoretical models and the role of configuration mixing is challenging for a number of theoretical reasons  \cite{34SiAlex}. The truncation of the model space cannot accommodate an explicit center of mass (CM) separation. Mixing can aggravate this problem and attempts to remove this contamination, such as using the Lawson technique, 
may impact the mixing and lead to noticeable uncertainty in energies. A careful treatment of mixing and associated with it the renormalization of the ground state, complicates the description of binding energies. The FSU interaction without mixing by initial design describes binding energies in a broad range of masses very well \cite{lubna2}. The introduction of mixing pushes levels apart and renormalizes the ground state in a significant way. This could be a reason for the shortcomings of the SDPF-U-MIX interaction mentioned in Ref. \cite{lica34Si}: there are problems with negative parity states and the introduction of mixing pushes the position of the $4^-$ state higher up in energy and further away from what is observed experimentally. In the present work no mixing between different configurations was allowed with the FSU interaction. Therefore, though the energy level predictions were quite satisfactory with the dominant $0\hbar \omega$ or $2\hbar \omega$ configurations, no calculated B(E2) values are shown as discussed in Ref. \cite{lubna2}. 

The negative parity states of $^{34}$Si were predicted with the FSU interaction quite successfully, as shown in Fig. \ref{fig:ThFsuAb}. The lowest $4^-$ level calculated at 4949 keV may correspond to the experimentally observed 4379 keV state with the occupancy dominated by one neutron excitation to the $f_{7/2}$ orbital. The experimentally observed 4970 keV state, whose spin could not be resolved from the $\gamma$-$\gamma$ angular correlation, can have theoretical counterparts at 4895 or 4948 keV with the spin-parity $3^-$ or $5^-$ respectively. If this were a $3^-$, 5041 keV would most likely be a $5^-$ depopulated by an E3 transition at 1716 keV which is 19 times stronger than the expected M1 transition at 663 keV. This does not support spin-parity $5^-$ assignment to the 5041 keV level. Hence we suggest $3^-$ to the 5041 keV level and $5^-$ to the 4970 keV state. The lowest-lying $3^-, 4^-$ and $5^-$ levels are all predicted to contain the dominant $(\nu d_{3/2})^{-1} \otimes (\nu f_{7/2})^{1}$ configuration as expected, though the energy predictions are better for the $3^-$ and the $5^-$ states as compared to that of the $4^-$ level. It is worth mentioning that the FSU shell-model interaction was utilized before to interpret the levels of $^{34}$Si in a review paper, Ref. \cite{34SiAlex}. In that work the intruder levels were predicted only for the neutron excitation from the $sd$ to the $fp$ shell.

The experimental states were also interpreted with the ab-initio approach by using the VS-IMSRG method \cite{Stroberg2019}. The calculations were performed with the 1.8/2.0 (EM) interaction \cite{Entem2003,Hebeler2011} expressed within the 15 major-shell harmonic oscillator (HO) space at $\hbar\omega=16$ MeV. The interaction is fitted to the data up to $A=4$ observables and can globally reproduce the ground-state energies~\cite{Simonis2017,Stroberg2021}, while underestimating the radii. For the three-nucleon matrix elements, one needs the additional $E_{\rm 3max}$ truncation, defined as the sum of the three-body HO quanta. Along with the recently introduced storage scheme \cite{Miyagi2022}, a sufficiently large $E_{\rm 3max}$ of 24 was used. The Hamiltonian was normal ordered with respect to the Hartree-Fock state obtained through the ensemble normal ordering \cite{Stroberg2017}, and the residual three-nucleon interaction was discarded. The standard $sd$ plus neutron $\{0f_{7/2}, 1p_{3/2}, 1p_{1/2}\}$ valence space was decoupled with the other space using a unitary transformation constructed in the VS-IMSRG framework with the two-body level approximation. Also, it was observed that the spurious center-of-mass modes due to the multi-shell valence space are stably removed with the Gl\"ockner-Lawson parameter $\beta=3$ (see Ref. \cite{vsimsrg} for further details). The VS-IMSRG calculations and subsequent diagonalizations were conducted with the imsrg++ \cite{imsrg++} and KSHELL \cite{Shimizu2019} codes, respectively.

From Fig. \ref{fig:ThFsuAb}, it can be seen that the VS-IMSRG  predicts the first two excited $0^+$ and $2^+$ states quite satisfactorily. Beyond that, there are not many theoretical candidates with reasonable energy match, especially for the $2^+$ levels populated by the isomeric states of $^{34}$Al; a total of three $2^+$ levels were predicted up-to 9 MeV. An energy shifting of 2 MeV of the predicted negative parity states make them good matches to the experimental states. The over prediction of the excitation energies were also reported in the earlier works. According to the discussions made in Refs \cite{Morris2018, Taniuchi}, the over prediction is most likely attributed to the two-body approximation employed in the VS-IMSRG.

\begin{table*}
\setlength{\tabcolsep}{0.6em}
\caption{The experimentally observed energy states are compared with their suggested theoretical counterparts, calculated with the FSU interaction, based on the excitation energies, $\gamma$-ray decay patterns, and the log$ft$ values. All the energy levels listed are in keV. The $1^+$ and $4^-$ states of the parent nucleus $^{34}$Al were calculated for the $1\hbar\omega$ and $0\hbar\omega$ excitations respectively. Because of the unmixed calculations for the particle-hole excitation for both in $^{34}$Al and $^{34}$Si, log$ft$ values with the FSU interaction were not calculated for the levels of $^{34}$Si with $0\hbar\omega$ configuration.}
\label{tab:thtab1}
\begin{tabular}{|c|c|c|c|c|}
\hline
$J^\pi$  & $\text E_{x(FSU)}$ & log$ft_{(FSU)}$ & $\text E_{x(Exp)}$ & log$ft$ \\[0.5ex] \hline
$0^+_1$ 	& 0   			&    				& 0	    																& 5.2(2) \cite{lica34Si}		\\ \hline	 						
$0^+_2$ 	& 2432		& 4.58		& 2719 															 	& 4.74(7) \cite{lica34Si}	\\ \hline	   						
$2^+_1$ 	& 3666		& 5.43		& 3326																& 5.4(2)	\cite{lica34Si}		 \\ \hline							
$2^+_2$ 	& 4359		& 5.65		& 4519																& 5.33(2)		\\ \hline													
$2^+_3$ 	& 5246		&	    			& 5349																&           			\\ \hline													
$0^+_3$ 	& 5426		& 5.35		&	    																	&		    			\\ \hline													
$2^+_4$ 	& 5493		& 10.02		& 5774																&		    		   \\ \hline
$2^+_5$ 	& 6675		& 5.12		& 6676																& 5.62(1)	   \\ \hline
$2^+_6$ 	& 6883		& 7.89		& 6764																&        			   \\ \hline
$0^+_4$ 	& 7122		& 4.60		& 7477																& 4.73(1)	   \\ \hline
$2^+_7$	& 7372		& 5.74		& 6873																&		    	       \\ \hline
$0^+_5$ 	& 7416		& 5.57		& 7107																& 4.99(1)	   \\ \hline
$2^+_8$ 	& 7693		& 6.83		&	    																	&	                  \\ \hline 
$3^-_1$		& 4466		& 4.43		& 4255																& 4.79(1)     \\ \hline
$3^-_2$		& 4895		& 5.71		& 5041																& 6.32(1)     \\ \hline
$5^-_1$		& 4948		& 5.36		& 4970																& 5.71(1)     \\ \hline
$4^-_1$		& 4949		& 4.54		& 4379																& 5.06(1)     \\ \hline
$3^-_3$		& 5748		& 5.05		& 6022																&  5.69(1)    \\ \hline
$4^-_2$		& 6369		& 5.67		& 6227																& 6.36(2)     \\ \hline
\end{tabular}
\end{table*}


The neutron occupancies of the $0f_{7/2}$ and $1p_{3/2}$ orbitals for the excited states calculated with the FSU interaction and the ab-initio method VS-IMSRG are also shown in Fig. \ref{fig:ThFsuAb}. It can be seen that up-to the second $2^+$ level, all the states have more $0f_{7/2}$ neutron occupancy than that of $1p_{3/2}$ predicted with both theoretical methods. The $0^+$ states at 7122 and 7416 keV calculated with the FSU interaction are dominated by the neutron $0f_{7/2}$ occupancy, whereas the level predicted at 7491 keV with the VS-IMSRG has almost equal neutron contribution to the $0f_{7/2}$ and $1p_{3/2}$ orbitals. We observed disagreements in the $fp$-shell occupancies between the results presented here and those reported in Ref. \cite{lica34Si}, our studies do not reproduce high occupancy of $f_{7/2}$ and $f_{5/2}$. Mixing of $\hbar\omega$ configurations within the FSU interaction does not have a significant enough impact on the occupancies to resolve this problem. The mixing is naturally present in VS-IMSRG results which also do not reproduce those reported in \cite{lica34Si}. The disagreements suggest significant difference in the Hamiltonians, possibly in some collective components which cannot be reconciled without comprehensive theoretical examinations of the Hamiltonians, and systematic studies of near-by nuclei. Such a theoretical study is beyond the scope of this work. 

For the negative parity states, the FSU calculation has restricted a maximum of one particle excitation within the $psdfp$ valence space. However, in all the cases shown for the VS-IMSRG approach, more than one neutron excitation from the $sd$ to the $fp$ shell has occurred. For the first $3^-$, $4^-$ and $5^-$ states, both the theoretical models predict that these levels are dominated by the neutron excitation to the $0f_{7/2}$ orbital. The third $3^-$ state calculated with FSU is dominated by the $1p_{3/2}$ occupation which arises from the neutron excitation across the $N=28$ shell gap, whereas this is still slightly dominated by the $0f_{7/2}$ neutron in the predictions of the VS-IMSRG. The first $4^-$ and $5^-$ levels predicted by both theories have the dominant $(\nu d_{3/2})^{-1} \otimes (\nu f_{7/2})^{1}$ configurations as expected. 

\begin{figure}
\begin{center}
\includegraphics[width=0.8\textwidth]{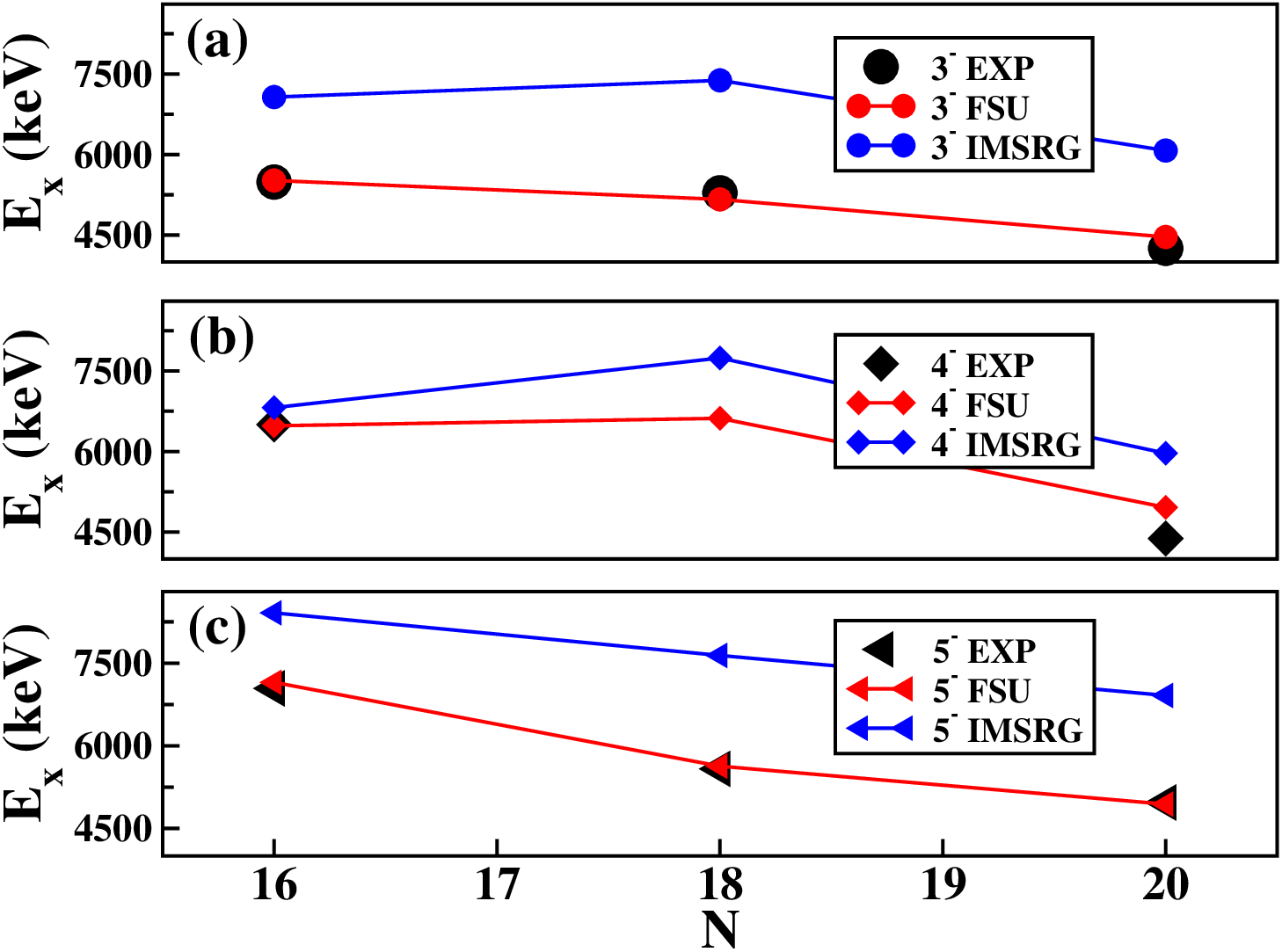}
\end{center}
\caption{Systematics of the negative parity states in the even-mass Silicon isotopes. The lowest-lying (a) $3^-$, (b) $4^-$ and (c) $5^-$ states, both experimental and predicted with  the FSU and the IMSRG-derived interactions, are shown here. The experimental levels, except for that of $^{34}$Si are compiled from Refs. \cite{datasheet30, datasheet32}.}
\label{fig:Si-Sys}    
\end{figure}

The systematics of the negative parity states in the even-mass Si isotopes are shown in Fig. \ref{fig:Si-Sys}. The simplest configurations for the first $3^-, \, 4^-$ and $5^-$ states in the $N\leq20$ isotopes are expected to be due to the uncoupling of two paired neutrons in the $sd$ shell and the promotion of one neutron to the $0f_{7/2}$ orbital with some rearrangements within the $sd$ shell. This is satisfied in the energy levels of $^{30}$Si, $^{32}$Si and $^{34}$Si, calculated with both the FSU and the IMSRG-derived interactions, with a gradual decrease in energies along with the increasing number of neutrons, indicating a reduction of the $N=20$ shell gap. Predictions made with the FSU interaction follow the experimental trend quite well. However, the IMSRG-derived interaction over-predicts the excitation energies for all the negative parity states. A constant energy shift for individual isotopes makes the predictions better with the latter interaction. Predicting the opposite-parity levels consistently with higher energy deviations in the even-Z even-N isotopes in this mass region has been discussed before \cite{lica34Si}. The agreement between the observed energies and those predicted by the FSU interaction is quite satisfactory for the Si isotopic chain with an RMS deviation of 224 keV which is a major improvement as highlighted in Ref. \cite{lubna2}. It will be quite informative to study the negative parity states of more exotic even-mass Si isotopes, as with higher neutron-to-proton ratio, the shell gap at $N=28$ will start to play a significant role. Those isotopes, which can be populated with the help of the new generation rare isotope beam facilities, will pose a significant challenge to the configuration interaction shell models as well as the models developed from more fundamental theories. 

\section{Summary}
An investigation on the structure of $^{34}$Si was carried out using $\beta$-delayed $\gamma$-ray spectroscopy with the GRIFFIN spectrometer at the TRIUMF-ISAC facility. Confirmation of all the previously observed energy levels and the $\gamma$-ray transitions validates the experimental approach. For the first time spin assignments in $^{34}$Si have been determined through $\gamma$-$\gamma$ angular correlation measurements. This analysis confirmed the location of the first excited $4^-$ and $5^-$ states. A second excited $0^+$ level was also confirmed with the same analysis. Coincident summing corrections have been performed for the $\gamma$-ray transitions observed and a controversial peak at 1053 keV was confirmed as a summed peak.

The configuration interaction model with the FSU Hamiltonian was utilized in order to interpret the experimental results in detail. The energies, log$ft$ values, and the expected dominant configurations of both positive and negative parity states were predicted quite successfully with the FSU shell-model interaction. The ab-initio method VS-IMSRG was also employed in order to interpret the observed levels of $^{34}$Si. Except of the first excited $0^+$ and $2^+$ states, the level energies were over-predicted with the interaction derived from the IMSRG method. This result demonstrates the need for future improvements in the models derived from more fundamental theories. 

The trends in the excitation energies of the known negative parity states of even-mass Si isotopes have been compared with those predicted with the interactions derived from the FSU Hamiltonian and the VS-IMSRG methods. The systematics clearly support the $N=20$ shell gap reduction with the increase in the neutron number and suggest the need for experimental data from more exotic isotopes. The systematics also shed light on the scope for future improvements in the theoretical models. 

\begin{acknowledgements}
We would like to thank the operations and beam delivery staff at TRIUMF for providing the radioactive beam. 
The GRIFFIN infrastructure has been funded jointly by the Canada Foundation for Innovation, the British Columbia Knowledge Development Fund (BCKDF), the Ontario Ministry of Research and Innovation (ON-MRI), TRIUMF and the University of Guelph. TRIUMF receives federal funding via a contribution agreement through the National Research Council Canada (NRC). This work was also supported by the U.S. Department of Energy Office of Science, Office of Nuclear Physics under Award Number DE-SC0020451(FRIB) and DE-SC0009883 (FSU) and National Science Foundation under grant No. NSF PHY-2012522 (FSU).
The Tennessee Technological University team is supported by the U. S. Department of Energy, office of Science through grant No. DE-SC0016988. 
Supporting GEANT4 simulations for the $\gamma$-$\gamma$ angular correlation analysis was supported by the U.S. Department of Energy (DOE) under contract no. DE-FG02-93ER40789 and used computing resources provided by the Open Science Grid \cite{OSG01,OSG02}, which is supported by the National Science Foundation Award Number 2030508.
This work was further supported by the European Research Council (ERC) under the European Union’s Horizon 2020 research and innovation programme (grant agreement No 101020842). This work was supported in part by the Deutsche Forschungsgemeinschaft (DFG, German Research Foundation) -- Project-ID 279384907 -- SFB 1245. T.M. would like to thank S.R. Stroberg for the imsrg++ code used to perform VS-IMSRG calculations. The VS-IMSRG calculations were performed with an allocation of computing resources on Cedar at WestGrid.
C.E.S. acknowledges support from the Canada Research Chairs program. This work was supported in part by the Natural Sciences and Engineering Research Council of Canada (NSERC).
\end{acknowledgements}


%
\end{document}